\documentclass[5p]{elsarticle}

\journal{Nuclear Instruments and Methods B}

\newcommand{\I}{\mathrm{i}}
\newcommand{\E}{\mathrm{e}}
\newcommand{\D}{\,\mathrm{d}}
\begin{document}

\begin{frontmatter}

\title{Kinetics of relativistic electrons passing through quasi--periodic fields}
\author[kiptaddress]{Eugene Bulyak\corref{corrauthor}}
\cortext[corrauthor]{Corresponding author}
\ead{bulyak@kipt.kharkov.ua}

\author[kiptaddress,KhGU]{Nikolay Shul'ga}
\address[kiptaddress]{NSC KIPT, 1 Academicheskaya str, Kharkov 61108 Ukraine}
\address[KhGU]{Karazin National University, 31, Kurchatov av., Kharkov 61108 Ukraine}

\begin{abstract}
We report a novel method for evaluating the energy spectrum of electrons emitting hard x-rays and gamma-rays in undulators and Compton sources. The method takes into account the quantum nature of recoils undergone by the electrons emitting high energy photons. The method is susceptible to evaluate a spectrum of electrons for the whole range of the emission rates per electron-pass through of the driving force, from much less than one emitted photon on average (Compton sources and short undulators) to many emitted photons (long undulators, relatively low-energy electrons). As shown in the former limiting case, the spectrum of electrons reflects the spectrum of emitted radiation whereas it is close to the Gaussian shape in the latter case. Limitation of coherency for the sources of high-energy electromagnetic radiation caused by recoils from emitted photons is also discussed.
\end{abstract}


\end{frontmatter}

\section{Introduction}

Bright intense beams of hard x- and gam\-ma rays are in increased demand for both fundamental research and for applications. The base method of generating such beams of photons is radiation from highly relativistic electrons passing through a periodic field, such as laser pulses in Compton sources, \cite{bulyak10,zen16}, undulators of XFELs \cite{xfel} or of the gamma sources of linear colliders, \cite{lc}. In a periodic field the energy of emitted quanta increases  with the squared energy of electrons and is inverse to the spatial period of the structure.
With the increase in the energy of photons, quantum effects connected with recoils come into play. These effects modify the energy spectrum of the electrons that in turn change the spectrum of the photons and the degree of their coherence.

Specific for the considered systems is that the maximal energy of the emitted radiation spectra is much smaller than electron's: For the undulators the maximal photon energy is $\sim 10^{-6}$ of the electron's (XFEL, \cite{xfel}) and $\sim 10^{-4}$ (LC gamma source, \cite{lc}). In the Compton gamma source the spectra will reach $\sim 0.02$ of the electron's, \cite{bulyak10,zen16}. The average number of photons emitted by the electron per pass through field -- ratio of the total energy of emitted radiation to the mean energy in the spectrum -- does not exceed a few hundreds for the undulator--based sources and less than unity for the Compton sources.

Modern sources of high--energy photons have nonuniform driving force along the trajectory of electrons: due to a particular envelope  of the laser pulse \cite{bulyak05} or its chirping in Compton sources, or implemented tapering in FELs.

In this paper, we present an analytic method for calculating the evolution of the spectrum of electrons passing through the periodic structures. The method is based upon the balance equation. The paper is organized as follows: In the first section we present a general method of evaluating the evolution of the spectrum of highly relativistic electrons spontaneously losing their energy while passing through a quasi--uniform field. The second section contains results of the application of the proposed method for evaluating the spectrum of the electrons emitting radiation in undulators and laser pulses (Compton inverse radiation). We also discuss the limitations imposed by the spectrum dilution upon the Compton sources and the undulator--based ones.

\section{Method}
\subsection{Problem setup and solution}
We will describe kinetics of the electrons that undergo distinct recoils
with the uni-dimensional balance equation analogous to that employed
for the distribution of the ionization losses, see \cite{landau44,shulga96}. This equation represents the conservation of `electron plus photon' energy.

Emission of photons causes degradation of the electron energy. The process of degradation is of stochastic nature and may be presented as a compound Markov process: the discontinuous photon emission, probability $\psi(z)$ ($z$ the axial coordinate), and the photon energy continuously distributed over the spectrum $w(\omega ;\gamma)$ ($\omega $ the reduced photon energy $\hbar\omega/mc^2$, $\gamma$ the Lorentz factor of the electrons, $\gamma =  E_\mathrm{e}/mc^2$). The spectrum is suggested  to be normalized: $\int_{-\infty}^\infty w(\omega ;\gamma)\D\omega = 1$. Here we use the infinite limits of integration, $\pm\infty$, supposing $w(\gamma,\omega <0)=0$. Further we omit the signs of infinite limits.

With changing the independent variable from $z$ to $x(z)$ (see \cite{khokonov04e}),
\[
x(z) = \int_0^z \psi (z) \D z\; ,
\]
equal to the mean number of  photons having been emitted from the beginning, an energy balance equation reads as
\begin{eqnarray} \label{eq:dinit}
\frac{\partial f(x,\gamma)}{\partial x}  &=& \int_{-\infty}^\infty \left[ w(\omega,\gamma + \omega) f(x,\gamma + \omega)- \right.\nonumber \\
 &&\left.  w(\omega,\gamma) f(x,\gamma )\right]\D\omega \, ,
\end{eqnarray}
where $f(x,\gamma)$ is a normalized to unity spectrum of electrons with $f(x,\gamma <1)=0$ and $f(0,\gamma)=f_0 (\gamma )$ being the initial distribution.

The key assumption to solve Eq.~(\ref{eq:dinit}) is that for the considered problem the spectrum of recoils is independent of the energy of the electron, $w(\omega ; \gamma+\omega)=w(\omega ;\gamma) = w(\omega )$, see \cite{landau44,shulga96}.

Under this simplification, the balance equation (\ref{eq:dinit}) can be reduced to
\begin{equation} \label{eq:basicsymb}
 \frac{\partial }{\partial x} f =  (f \star  w)(\gamma )  - f \, ,
\end{equation}
where $\star $ indicates the cross--correlation operation,
\[
(f \star  w)(\gamma ) = \int f(x,\gamma +\omega) w(\omega) \D\omega \, .
\]

The Fourier transform of the balance equation is
\begin{equation} \label{eq:dfourier}
\frac{\partial }{\partial x} \hat{f} = \hat{f} \, \check{w}   - \hat{f} = \hat{f} \, (\check{w} -1 ) \, ,
\end{equation}
where $\hat{f}$ is the Fourier transform of the electron spectrum, $\mathcal{F}[f]$, and $\check{w} = \mathcal{F}^{-1}[w]$ the inverse Fourier transform of the recoils spectrum. We used  the fact that the Fourier transform of the cross--correlation is the product of the direct and the inverse transforms.

The characteristic function, completely defined evolution of the spectrum of electrons, is a solution to (\ref{eq:dfourier}):
\begin{equation} \label{eq:foursol}
\hat{f} = \E^{-x}\E^{x\check{w}}\hat{f_0}\, ,
\end{equation}
with $\hat{f_0}$ being the Fourier transform of the initial spectrum of electrons, $f_0 (\gamma) \equiv f(x=0,\gamma)$.

\subsection{Analysis}
Development of the second exponent in (\ref{eq:foursol}), $\E^{x\check{w}}$, into  Taylor series and then the inverse Fourier transform yields:
\begin{eqnarray}
\hat{f} &=& \sum_{n=0}^\infty \frac{\E^{-x} x^n}{n!}\,\hat{f_0}{\check{w}}^n\, ; \label{eq:fourexp} \\
f(x,\gamma) &=& \sum_{n=0}^\infty \frac{\E^{-x} x^n}{n!}\,F_n(\gamma)\, ,
\label{eq:reexp}
\end{eqnarray}
where $F_n(\gamma)$ is $n$--state density distribution -- the spectrum of electron having emitted exactly $n$ photons.

The states may be calculated in two different equivalent ways, as (i) the inverse Fourier transform or (ii) the recursive solution of the Chapman--Kolmogorov equations, \cite{kolchuzhkin02}:
\begin{eqnarray}
F_n^{(i)}(\gamma) &=& \mathcal{F}^{-1}\left[ \hat{f_0}{\check{w}}^n\right]\, ; \label{eq:nstatef} \\
F_n^{(ii)}(\gamma) &=& \int F_{n-1}(\gamma+\omega) w(\omega)\D\omega\, , \label{eq:nstateck}
\end{eqnarray}
with $F_0(\gamma) \equiv f_0(\gamma)$ being the initial spectrum.

Thus, the evolution of the electron spectrum in quasi--pe\-ri\-o\-dic fields may be described as superposition of $n$-states with the Poisson mass of parameter $x\ge 0$ equal to the mean number of emitted photons. From Eq.(\ref{eq:reexp}) it immediately follows evolution of the mean energy
\begin{equation}\label{eq:meanener}
\overline{\gamma} = \left.\overline{\gamma}\right|_0 -x\, \overline{\omega}\; ,
\end{equation}
Also this presentation allows to derive  $m$-th central moment of electron spectrum that reads:
\begin{equation} \label{eq:moments}
\overline{\left(\gamma-\overline{\gamma}\right)^m} = \left.\overline{\left(\gamma-\overline{\gamma}\right)^m}\right|_0 + (-1)^m x\, \overline{\omega^m}\; ,
\end{equation}
where `overline' sign indicates the ensemble average, the initial magnitude of the moment subscribed by `0' sign.

It should be pointed out that $m$-th \emph{central moment} is proportional to the average number of recoils, $x$, and to $m$-th \emph{raw moment} of the photon spectrum. Negative third moment -- skewness -- specifies asymmetry of the electron spectrum, where the `tail' expands towards lower energies. Particularly, the negative skewness indicates that the mode of electron spectrum (maximum in spectrum) is at the higher energy than the mean:
\[
\max f(x,\gamma)=f(x,\gamma_\mathrm{mode}) \to \gamma_\mathrm{mode} \approx \overline{\gamma} + \frac{\overline{\omega^3}}{2\overline{\omega^2}}\; .
\]

The considered stochastic process being of non-dif\-fu\-sive nature, see \cite{feller49}, allows for diffusive description with the two first moments for big number of recoils, $x\gg 1$, normalized (Pearson's) skewness tend to zero as $1/\sqrt{x}$.

\section{Dipole radiation}
Compton inverse radiation as well as from weak undulators is the dipole radiation (see \cite{luchini90}) with a rather simple spectral shape.

A practical example of the radiating system is that of emitting the first harmonic which takes place in the Compton sources and weak undulators. The recoil spectrum reads, see e.g. \cite{bulyak15}:
\begin{equation} \label{eq:dipole}
w(\omega) = \frac{3}{2\omega_\mathrm{m}}  \left[1 -\frac{2 \omega}{\omega_\mathrm{m}} \left( 1-\frac{\omega }{\omega_\mathrm{m}} \right) \right]  \Pi \left( \frac{\omega}{\omega_\mathrm{m}} - \frac{1}{2} \right)\,,
\end{equation}
where $\Pi(x)$ is the `rect' function (equal 1 within $|x| < 1/2$ and zero beyond), $\omega_\mathrm{m}$ is the maximal photon energy in the first harmonic.

Moments of the recoil, $\overline{\omega } = \omega_\mathrm{m}/2$,
$\overline{\omega^2 } = \frac{7}{20}\omega_\mathrm{m}^2$ and $\overline{\omega^3} =\frac{11}{40}\omega_\mathrm{m}^3$, produce evolution of the centered moments of the electron spectrum:
\begin{eqnarray*}
\overline{\gamma }(x) &=&  \overline{\gamma_0} - x\, \omega_\mathrm{m}/2\, , \\
\mathrm{Var}\left[\gamma\right] (x) &\equiv &\overline{\left(\gamma-\overline{\gamma}\right)^2} = \mathrm{Var}\left[\gamma_0\right] + \frac{7}{20}x\,\omega_\mathrm{m}^2  \, ,\\
\mathrm{Sk}\left[\gamma\right] (x) &\equiv & \overline{\left(\gamma-\overline{\gamma}\right)^3}=\mathrm{Sk}\left[\gamma_0\right] - \frac{11}{40}x\,\omega_\mathrm{m}^3\, .
\end{eqnarray*}

For the initial delta--like spectrum  $f_0(\gamma )=\delta(\gamma - \gamma_0)$ of electrons the Fourier transform for $F_n$ read
\begin{equation}
\hat{F}_n(s)=
 e^{-2 \I \pi  s (\gamma_0 - n \overline{\omega} )}\times %&&\\
\end{equation}
\[
\left[\frac{3}{16}\,\frac{\left(4 \pi
   ^2 s^2 \overline{\omega} ^2-1\right) \sin (2 \pi  s \overline{\omega})+2 \pi  s \overline{\omega}  \cos (2 \pi  s \overline{\omega} )}{(\pi s\overline{\omega})^3
   }\right]^n .
\]

\begin{figure}[htb]
   \includegraphics[width=\columnwidth]{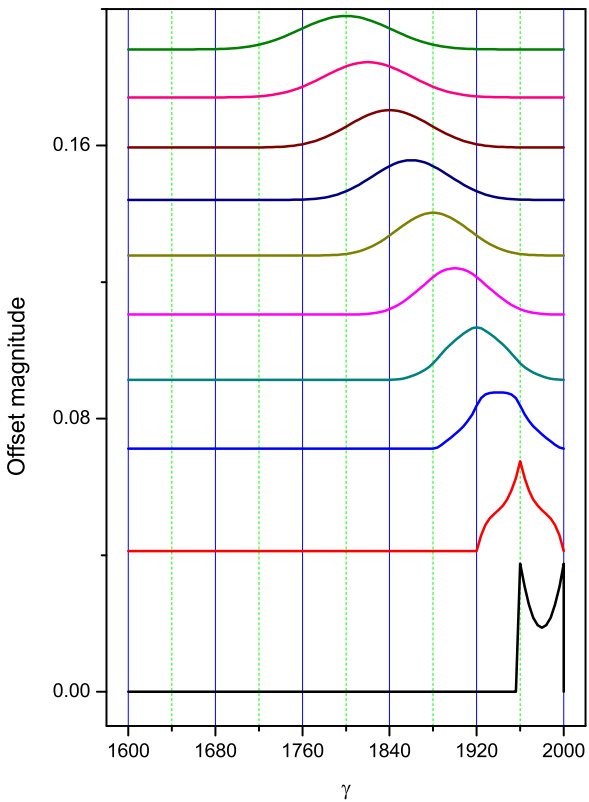}
   \caption{First ten specific spectra of electrons (from bottom to top). $\gamma_0 = 2000$, $\omega_\mathrm{m}=40$.}
   \label{fig:anal10}
\end{figure}

The first ten states for the initial delta distribution are presented in Fig.~\ref{fig:anal10} (cf. numerical results in \cite{kolchuzhkin02}).

As it can be seen, with an increase in the number of scattered quanta the shape of the specific state is losing its individuality and is converging to the Gaussian, in accordance with the Central limit theorem.

\begin{figure*}[htb]
\centering
   \includegraphics[width=0.8\columnwidth]{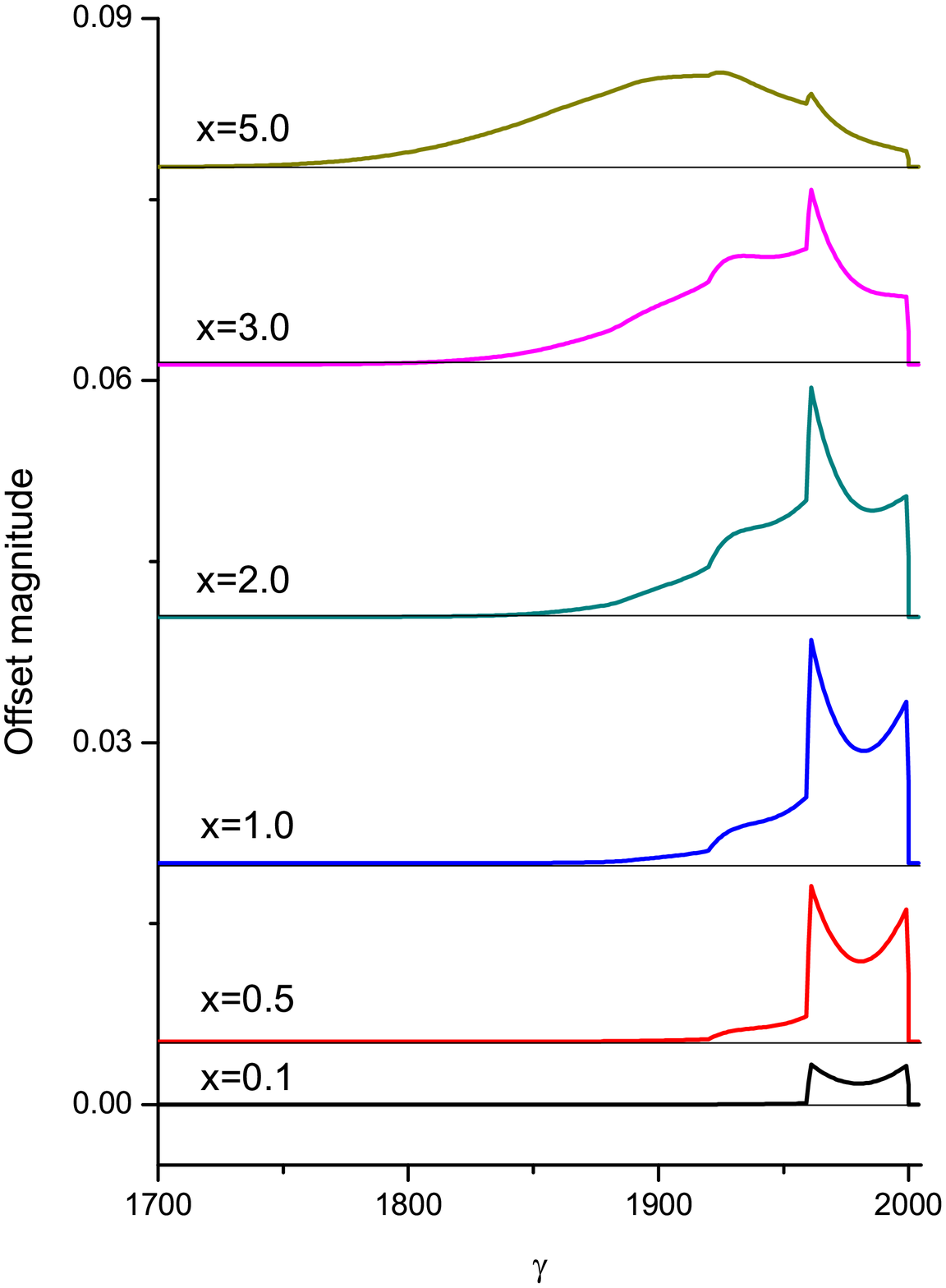}
   \includegraphics[width=0.8\columnwidth]{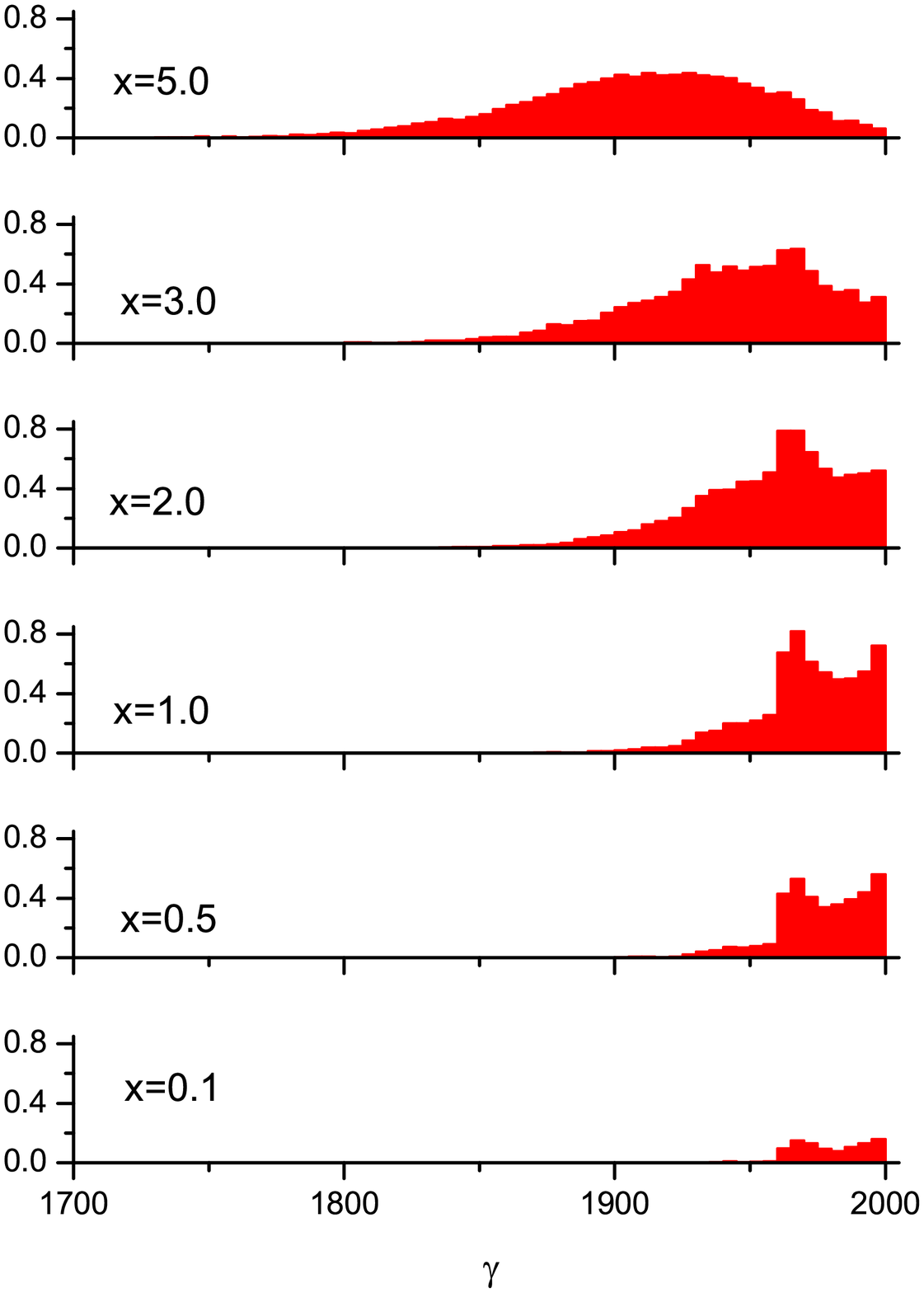}
   \caption{Aggregate spectra at different average numbers of recoils $x$. Left panel -- calculation, right one -- simulation.}
   \label{fig:tenstates}
\end{figure*}

The aggregate spectra composed from the first ten states with the Poisson mass for the different average number of emitted photons are presented in Fig.~\ref{fig:tenstates} (left panel), the initial delta distribution $F_0(\gamma)$ is not included in the spectra. For comparison, we did Monte Carlo simulations for the Compton source with the same $\gamma_0$ and $\omega_\mathrm{m}$,  Fig.~\ref{fig:tenstates} (right panel).

Both the analytic and the simulated spectra display similar behaviour at the beginning of the photon emission. These spectra radically differ from the normal distribution at the small average number of emitted quanta and tend to the normal with the increase of $x$. (The spectra are in agrement with those calculated by other authors, see \cite{kolchuzhkin02,petrillo13a}.)

Spreading out of the electron spectra imposes some limitations upon the undulator based and Compton sources of high energy radiation. Requirement for the coherency of radiation from the electrons traversing the FEL undulator imposes restriction upon the width of the electron spectrum \cite{pellegrini16}:
\begin{equation} \label{eq:disGam}
\frac{\Delta\gamma}{\gamma }\approx \frac{\sqrt{\mathrm{Var}[\gamma]}}{\gamma } = \frac{1}{2 N_\mathrm{u}}\, ,
\end{equation}
where $N_\mathrm{u}$ is the number of undulator periods.

Taking into account the number of periods needed to emit one photon, see
\cite{bulyak15}, $\nu_1 = 3/(4\pi \alpha K^2)$ ($K$ is the undulator parameter, $\alpha$ the fine structure constant), and contribution to the variance from one recoil,
$\overline{\omega^2}\approx  7/20\times\omega_\mathrm{m}^2$, we obtain a maximum number of the undulator periods for which the recoils induce energy spread equal to (\ref{eq:disGam}):
\begin{equation} \label{eq:xlimit}
N_\mathrm{coh} \approx 2\left( \frac{\lambda_\mathrm{u}}{K\gamma\lambda_\mathrm{C}} \right)^{2/3} \, ,
\end{equation}
where $\lambda_\mathrm{u}$ is the undulator spatial period, $\lambda_\mathrm{C}$ is the Compton wavelength for the electron.

The low--energy tail in the electron spectrum that occurred at a small number of emitted photons -- in Compton sources -- can cause beam losses due to limited energy acceptance in the Compton rings \cite{bulyak10}, or decreased efficiency of the energy recovery in sources based on Energy Recovery Linac (ERL) \cite{gruner02}.

\section{Conclusions}
We have presented a method for evaluating the evolution of the spectrum of relativistic electrons passing through a periodic field in the hard x--ray and gamma--ray sources. As is rigorously shown, the spectrum may be presented as the Poisson--mass sum of the $n$-states each of which corresponds to exactly $n$ recoils undergone by the electron. The Poisson distribution parameter, equal to the average number of recoils, represents evolution; the $n$-states are determined by the initial spectrum and the spectrum of the recoils due to the emission of the high--energy photons. The variance (the second central moment) of the electron spectrum is determined by the dispersion (the second raw moment) of the spectrum of recoils. The same principle is valid for the higher moments as well.
This method provides specific spectrum at the small average number of recoil events that corresponds to the Compton sources and the beginning section of the undulator--based sources. With an increase in the average number of emitted photons, the aggregate spectra tends to the Gaussian, and the diffusion approximation to the kinetic equation becomes valid. The variance of the electron spectrum is determined by the dispersion of the spectrum of recoils. Due to the negative skewness of the electron spectrum, maximum density of the electrons is shifted to a higher energy form the mean energy.

\subsection*{Acknowledgments}
This work is partially supported by the Ministry of Education and Science of Ukraine,
project No 1-13-15.

\section*{References}
\providecommand{\noopsort}[1]{}\providecommand{\singleletter}[1]{#1}%

\end{document}